\documentstyle[epsf]{article}

\begin{document}

\renewcommand\baselinestretch{1.4}
\large\normalsize

\title{Square--lattice $s=\frac{1}{2}$ $XY$ model
       and the Jordan--Wigner fermions:
       \protect\\
       The ground--state and thermodynamic properties}

\author{Oleg Derzhko$^{\dagger,\ddagger}$,
        Taras Verkholyak$^{\dagger}$,
        Reimar Schmidt$^{\star}$ 
        and
        Johannes Richter$^{\star}$\\
{\small {$^{\dagger}$Institute for Condensed Matter Physics,}}\\
{\small {1 Svientsitskii Street, L'viv--11, 79011, Ukraine}}\\
{\small {$^{\ddagger}$Chair of Theoretical Physics,
Ivan Franko National University of L'viv,}}\\
{\small {12 Drahomanov Street, L'viv--5, 79005, Ukraine}}\\
{\small {$^{\star}$Institut f\"ur Theoretische Physik,
Universit\"at Magdeburg,}}\\
{\small {P.O. Box 4120, D--39016 Magdeburg, Germany}}}

\date{\today}

\maketitle

\begin{abstract}

Using the 2D Jordan--Wigner transformation
we reformulate
the square--lattice
$s=\frac{1}{2}$ $XY$ ($XZ$) model
in terms of noninteracting spinless fermions
and examine
the ground--state and thermodynamic properties
of this spin system.
We consider the model
with two types of anisotropy:
the spatial anisotropy
interpolating between 2D and 1D lattices
and
the anisotropy of the exchange interaction
interpolating between isotropic $XY$ and Ising interactions.
We compare the obtained (approximate) results
with exact ones
(1D limit, square--lattice Ising model)
and other approximate ones
(linear spin--wave theory
and exact diagonalization data
for finite lattices of up to $N=36$ sites
supplemented by finite--size scaling).
We discuss the ground--state and thermodynamic properties
in dependence on 
the spatial and exchange interaction anisotropies.
We pay special attention to
the quantum phase transition
driven by the exchange interaction anisotropy
as well as to
the appearance/disappearance of the zero--temperature magnetization
in the quasi--1D limit.

\end{abstract}

\vspace{5mm}


\vspace{5mm}


\vspace{5mm}

\noindent
{\bf {Postal address:}}\\
Dr. Oleg Derzhko (corresponding author)\\
Institute for Condensed Matter Physics,\\
1 Svientsitskii Street, L'viv--11, 79011, Ukraine\\
tel/fax: 380 322 761978\\
email: derzhko@icmp.lviv.ua

\clearpage

\renewcommand\baselinestretch{1.85}
\large\normalsize

\section{Introduction}

\setcounter{equation}{0}

The $s=\frac{1}{2}$ Heisenberg model on a square lattice
is the well--known basic model in the quantum theory of magnetism
which became especially attractive
after the discovery of high--temperature superconductors
\cite{001,002}.
Much more materials which can be viewed
as realizations of the 2D $s=\frac{1}{2}$  Heisenberg model
are known at present time.
Many analytical and numerical studies
analyzing thermodynamics, spin correlations and their dynamics
for the 2D $s=\frac{1}{2}$ Heisenberg model
were performed during the last fifteen years.
One of the (approximate) analytical approaches
is based on
the 2D Jordan--Wigner fermionization.
An extension of the 1D Jordan--Wigner transformation
for higher dimensions was suggested in the late 80s \cite{004}
(see also \cite{005,006,007}).
Such kind of approximate treatment
was applied to the square--lattice quantum spin models in several papers
\cite{004,005,006,008}
(for a brief review see \cite{009}).
Recently,
the 2D Jordan--Wigner mapping has been used
for the calculation of the magnetization curves
of several 2D spin systems \cite{010}.
The reported studies \cite{004,005,006,008,010}
deal with the $s=\frac{1}{2}$
isotropic Heisenberg
or the $s=\frac{1}{2}$ isotropic $XY$ models
and the effects
of anisotropic $XY$ exchange interaction
were not considered.
Another intriguing problem in the theory of magnetism
is the influence
of the spatial anisotropy
on the properties of quantum spin models
between the limiting cases
of a system of noninteracting chains
and of a spatially isotropic 2D system.
The issue of disappearing of long--range order in 1D limit
although discussed by several authors
(see, for example, the papers \cite{011,012,013,014}
on the crossover from one to two dimensions 
for the $s=\frac{1}{2}$ Heisenberg model)
remains ambiguous.

In this paper
we consider
the $s=\frac{1}{2}$ anisotropic $XY$ model
on a spatially anisotropic square lattice
using the 2D Jordan--Wigner fermionization approach
in order
1) to examine the effects of exchange interaction anisotropy
and the effects of spatial anisotropy
on the ground--state and thermodynamic properties
of the spin model
and
2) to gain an understanding
how well this technique works
comparing the approximate results with the exact ones
(1D limit,
the Onsager solution for the square--lattice Ising model).
To examine the quality of the 2D Jordan--Wigner fermionization scheme
in the cases for which exact results are not available
we perform linear spin--wave theory calculations
and finite--size scaling analysis
of exact diagonalization data.
The 2D Jordan--Wigner fermionization may be used
for a study of the dynamic properties of 2D quantum spin systems
\cite{015}
and a first test of this approach for calculating the static quantities
is strongly desirable.

There has been a great deal of theoretical work
on the 2D $s=\frac{1}{2}$ $XY$ model
\cite{016,017,018,019,020,021,022,023,024,025,026,027}.
We use the results reported in these papers
in our study for comparison in due course.
Thus, in Ref. \cite{018}
it was shown how the spin--wave theory can be applied
to the quantum $XY$ Hamiltonians.
Already the very early studies
on square--lattice $s=\frac{1}{2}$ isotropic $XY$ model
indicated
the long--range order in the ground state.
Later the existence of long--range order in the ground state
of quantum $XY$ model on a hypercubic lattice
in two and higher dimensions was proved rigorously \cite{019}.
The exact diagonalization study for the isotropic $XY$ model
on finite square lattices of up to $6\times 6$ sites
supplemented by finite--size scaling
was reported in Ref. \cite{022}.
The high precision quantum Monte Carlo results
presented in Ref. \cite{023}
concern with both the properties at zero and nonzero temperatures.
The zero--temperature properties
of the square--lattice
$s=\frac{1}{2}$ anisotropic $XY$ model
were studied via spin--wave theory
and via series expansions around the Ising limit \cite{024}.
The papers \cite{025,026}
contain the zero--temperature results
of the coupled cluster method
and the correlated basis function method applied 
on the square--lattice $s=\frac{1}{2}$ anisotropic $XY$ model,
respectively.
The Green's--function approach was used for
the calculation of the temperature dependences
of the susceptibilities and specific heat 
of the square--lattice $s=\frac{1}{2}$ isotropic $XY$ model  
in Ref. \cite{027}.
It should be remarked
that the mentioned studies refer to the spatially isotropic lattices.

This paper is organized as follows.
In Section 2 we present the conventional linear spin--wave theory 
for the square--lattice $s=\frac{1}{2}$ $XY$ model.
This consideration
is a straightforward generalization of Ref. \cite{018}
for the anisotropic case
and is given here for easy references.
In Section 3 we present the 2D Jordan--Wigner fermionization treatment
of the $XY$ and $XZ$ Hamiltonians.
The results derived in this Section are parallel to those
obtained on the basis of the spin--wave theory
and resemble strongly some outcomes
of the coupled cluster method
and the correlated basis function method \cite{025,026}.
In Section 4 we present
the finite--size scaling analysis of the exact diagonalization data.
In Section 5
we discuss the effects of anisotropies
on the ground--state and thermodynamic properties
of the square--lattice $s=\frac{1}{2}$ $XY$ model.
We compare the results of different approaches
focusing on the validity of the approximate treatment
based on the 2D Jordan--Wigner fermionization.
Finally, we summarize our findings in Section 6.
Some preliminary results of this study
were announced in the conference papers \cite{032}.

\section{The model.
Linear spin--wave theory}

\setcounter{equation}{0}

We consider
a model consisting of $N=N_x N_y$
($N_x=N_y=\sqrt{N}\to\infty$)
spins $\frac{1}{2}$
on a spatially anisotropic square lattice
governed by the anisotropic $XY$ Hamiltonian
\begin{eqnarray}
\label{2.01}
H
=\sum_{i=0}^{N_x-1}\sum_{j=0}^{N_y-1}
\left(
J\left((1+\gamma)s^x_{i,j} s^x_{i+1,j}
+(1-\gamma)s^y_{i,j} s^y_{i+1,j}\right)
\right.
\nonumber\\
\left.
+
J_{\perp}\left((1+\gamma)s^x_{i,j} s^x_{i,j+1}
+(1-\gamma)s^y_{i,j} s^y_{i,j+1}\right)
\right)
\nonumber\\
=\frac{1}{2}\sum_{i=0}^{N_x-1}\sum_{j=0}^{N_y-1}
\left(
J\left(s^+_{i,j} s^-_{i+1,j}+s^-_{i,j} s^+_{i+1,j}
+\gamma
\left(s^+_{i,j} s^+_{i+1,j}+s^-_{i,j} s^-_{i+1,j}\right)
\right)
\right.
\nonumber\\
\left.
+
J_{\perp}\left(s^+_{i,j} s^-_{i,j+1}+s^-_{i,j} s^+_{i,j+1}
+\gamma
\left(s^+_{i,j} s^+_{i,j+1}+s^-_{i,j} s^-_{i,j+1}\right)\right)
\right).
\end{eqnarray}
Here $J$ and $J_{\perp}$
($=R J$)
are the exchange interaction strengths
between neighbouring sites
in a row (horizontal direction)
and a column (vertical direction),
respectively,
and the parameter $\gamma$ controls the anisotropy of exchange interaction.
Since the considered lattice is a bipartite one,
the cases of ferromagnetic and antiferromagnetic signs of exchange interactions
are related to each other
by a simple spin rotation.
Therefore, in what follows
we assume without loss of generality the ferromagnetic sign
of the exchange interactions $J$ and $J_{\perp}$,
i.e., $J,J_{\perp}<0$.
We can recover the 1D limit
putting either $J_{\perp}=0$ or $J=0$
arriving at a system of independent chains
running in horizontal or vertical direction,
respectively.
The case $\gamma=0$ corresponds to the isotropic $XY$ interaction,
whereas the case $\gamma=1$
(or $\gamma=-1$)
corresponds to the Ising interaction.
We are interested
in the ground--state and thermodynamic properties
of spin model (\ref{2.01}).

We begin with the linear spin--wave theory
following the scheme developed in \cite{018}.
The conventional spin--wave analysis
starts from a ferromagnetic state fully polarized in $z$ direction 
\cite{033}.
We have to deal with the $XZ$ model 
rather than the $XY$ model.
The former Hamiltonian arises from the latter one
after the unitary transformation
$s^x\to -s^z$,
$s^y\to  s^x$,
$s^z\to -s^y$,
and hence exhibits the same thermodynamics.
(Evidently,
within approximate approaches
in which the $x$, $y$ spin components
are treated differently than the $z$ spin components
we may expect 
different results for the $XY$ and $XZ$ models.)
The performed rotation is also employed
within the coupled cluster method \cite{025}
or correlated basis function method  \cite{026}
applied to the quantum $XY$ model
(but not for exact diagonalization computations).
Thus, we consider
\begin{eqnarray}
\label{2.02}
H
=\sum_{i=0}^{N_x-1}\sum_{j=0}^{N_y-1}
\left(
J\left((1-\gamma)s^x_{i,j} s^x_{i+1,j}
+(1+\gamma)s^z_{i,j} s^z_{i+1,j}\right)
\right.
\nonumber\\
\left.
+
J_{\perp}\left((1-\gamma)s^x_{i,j} s^x_{i,j+1}
+(1+\gamma)s^z_{i,j} s^z_{i,j+1}\right)
\right)
\nonumber\\
=\sum_{i=0}^{N_x-1}\sum_{j=0}^{N_y-1}
\left(
J\frac{1-\gamma}{4}
\left(s^+_{i,j} s^+_{i+1,j}+s^+_{i,j} s^-_{i+1,j}
+s^-_{i,j} s^+_{i+1,j}+s^-_{i,j} s^-_{i+1,j}\right)
\right.
\nonumber\\
\left.
+J\left(1+\gamma\right)
\left(s^+_{i,j} s^-_{i,j}-\frac{1}{2}\right)
\left(s^+_{i+1,j} s^-_{i+1,j}-\frac{1}{2}\right)
\right.
\nonumber\\
\left.
+
J_{\perp}\frac{1-\gamma}{4}
\left(s^+_{i,j} s^+_{i,j+1}+s^+_{i,j} s^-_{i,j+1}
+s^-_{i,j} s^+_{i,j+1}+s^-_{i,j} s^-_{i,j+1}\right)
\right.
\nonumber\\
\left.
+J_{\perp}\left(1+\gamma\right)
\left(s^+_{i,j} s^-_{i,j}-\frac{1}{2}\right)
\left(s^+_{i,j+1} s^-_{i,j+1}-\frac{1}{2}\right)
\right).
\end{eqnarray}

Employing the Holstein--Primakoff transformation
we arrive at
the following Hamiltonian of the linear spin--wave theory
\begin{eqnarray}
\label{2.03}
H
=\sum_{i=0}^{N_x-1}\sum_{j=0}^{N_y-1}
\left(
J
\frac{1-\gamma}{4}
\left(a^+_{i,j}a^+_{i+1,j}+a^+_{i,j}a_{i+1,j}
+a_{i,j}a^+_{i+1,j}+a_{i,j}a_{i+1,j}\right)
\right.
\nonumber\\
\left.
+J(1+\gamma)
\left(\frac{1}{4}
-\frac{1}{2}a^+_{i,j}a_{i,j}
-\frac{1}{2}a^+_{i+1,j}a_{i+1,j}\right)
\right.
\nonumber\\
\left.
+J_{\perp}
\frac{1-\gamma}{4}
\left(a^+_{i,j}a^+_{i,j+1}+a^+_{i,j}a_{i,j+1}
+a_{i,j}a^+_{i,j+1}+a_{i,j}a_{i,j+1}\right)
\right.
\nonumber\\
\left.
+J_{\perp}(1+\gamma)
\left(\frac{1}{4}
-\frac{1}{2}a^+_{i,j}a_{i,j}
-\frac{1}{2}a^+_{i,j+1}a_{i,j+1}\right)
\right),
\end{eqnarray}
where the operators $a^+_{i,j}$, $a_{i,j}$
obey the Bose commutation rules.
After performing the Fourier transformation
\begin{eqnarray}
\label{2.04}
a_{i,j}=\frac{1}{\sqrt{N_xN_y}}
\sum_{k_x,k_y}
{\mbox{e}}^{{\mbox{i}}
\left(k_xi+k_yj\right)}
a_{k_x,k_y},
\;\;\;
a^+_{i,j}=\frac{1}{\sqrt{N_xN_y}}
\sum_{k_x,k_y}
{\mbox{e}}^{-{\mbox{i}}
\left(k_xi+k_yj\right)}
a^+_{k_x,k_y}
\end{eqnarray}
with
$k_x=\frac{2\pi}{N_x}n_x$,
$n_x=-\frac{N_x}{2},-\frac{N_x}{2}+1,\ldots,\frac{N_x}{2}-1$,
$k_y=\frac{2\pi}{N_y}n_y$,
$n_y=-\frac{N_y}{2},-\frac{N_y}{2}+1,\ldots,\frac{N_y}{2}-1$
($N_x$, $N_y$ are even)
we come to the following quadratic form
\begin{eqnarray}
\label{2.05}
H=
\frac{1}{4}\sum_{{\bf{k}}}
\left(
\left(
\left(1-\gamma\right)J_{\bf{k}}
-2\left(1+\gamma\right)\left(J+J_{\perp}\right)
\right)
\left(a^+_{{\bf{k}}}a_{{\bf{k}}}
+a_{-{\bf{k}}}a^+_{-{\bf{k}}}\right)
+\left(1-\gamma\right)J_{\bf{k}}
\left(a^+_{{\bf{k}}}a^+_{-{\bf{k}}}
+a_{{\bf{k}}}a_{-{\bf{k}}}\right)
\right)
\nonumber\\
+\frac{3}{4}N(1+\gamma)\left(J+J_{\perp}\right),
\\
J_{\bf{k}}
=J\cos k_x + J_{\perp}\cos k_y.
\nonumber
\end{eqnarray}
This quadratic form (\ref{2.05}) can be diagonalized
by the Bogolyubov transformation
\begin{eqnarray}
\label{2.06}
a_{{\bf{k}}}
=\left(\cosh u_{\bf{k}}\right)b_{{\bf{k}}}
+\left(\sinh u_{\bf{k}}\right)b^+_{-{\bf{k}}},
\nonumber\\
a^+_{-{\bf{k}}}
=\left(\sinh u_{\bf k}\right)b_{\bf k}
+\left(\cosh u_{\bf k}\right)b^+_{-{\bf k}},
\end{eqnarray}
where the real function $u_{\bf k}=u_{-\bf k}$
is given by the equation
\begin{eqnarray}
\label{2.07}
\tanh\left(2u_{\bf k}\right)
=\frac{\left(1-\gamma\right)J_{\bf k}}
{2(1+\gamma)\left(J+J_{\perp}\right)
-\left(1-\gamma\right)J_{\bf k}}.
\end{eqnarray}
As a result we get
\begin{eqnarray}
\label{2.08}
H
=\sum_{\bf k}\Omega_{\bf k}
\left(b^+_{\bf k}b_{\bf k}+\frac{1}{2}\right)
+\frac{3}{4}N(1+\gamma)(J+J_{\perp}),
\\
\Omega_{\bf k}
=\sqrt{\left(1+\gamma\right)^2\left( J+J_{\perp}\right)^2
-\left(1-\gamma^2\right)\left( J+J_{\perp}\right)J_{\bf k}}.
\nonumber
\end{eqnarray}
In the special case $\gamma=0$, $J=J_{\perp}$
Eq. (\ref{2.08}) reproduces the result derived in Ref. \cite{018}.

Now we can calculate in the usual way the ground--state energy per site
\begin{eqnarray}
\label{2.09}
e_0
=\frac{1}{2}
\int_{-\pi}^{\pi}\frac{{\mbox{d}}k_x}{2\pi}
\int_{-\pi}^{\pi}\frac{{\mbox{d}}k_y}{2\pi}
\Omega_{\bf k}
+\frac{3}{4}(1+\gamma)(J+J_{\perp})
\end{eqnarray}
and the zero--temperature ($z$) magnetization per site
\begin{eqnarray}
\label{2.10}
m^z
=-\frac{1}{2}
\int\frac{{\mbox{d}}{\bf {k}}}{\left(2\pi\right)^2}
\frac{1}{\sqrt{1-\tanh^2\left(2u_{\bf k}\right)}}
+1
\end{eqnarray}
(keeping in mind that $s^z=\frac{1}{2}-a^+a$).

Let us note that the approximate treatment
which starts from Eq. (\ref{2.03})
has destroyed the symmetry with respect to
the change $\gamma$ to $-\gamma$.
Moreover,
the consideration is valid only for $\gamma\ge 0$
otherwise
there exist such values of ${\bf{k}}$
at which  the r.h.s. of (\ref{2.07}) for
$\tanh\left(2u_{\bf k}\right)$
exceeds 1.
Evidently,
while getting (\ref{2.03})
we have presumed the  long--range order in $z$ direction in the spin space
and this assumption becomes wrong if $\gamma<0$.

\section{2D Jordan--Wigner fermions: $XY$ vs. $XZ$}

\setcounter{equation}{0}

An alternative approach
to calculate the thermodynamic quantities
of the square--lattice $s=\frac{1}{2}$ $XY$ model
consists in 
the 2D Jordan--Wigner transformation 
reformulating the problem in fermionic language.
First we introduce
the annihilation and creation operators of spinless fermions
\begin{eqnarray}
\label{3.01}
s_{i,j}^-
={\mbox{e}}^{{\mbox{i}}\alpha_{i,j}}d_{i,j},
\;\;\;
s^+_{i,j}
={\mbox{e}}^{-{\mbox{i}}\alpha_{i,j}}d_{i,j}^+,
\nonumber\\
\alpha_{i,j}
=\sum_{f(\ne i)}
\sum_{g(\ne j)}
{\mbox{Im}}\ln\left(f-i+{\mbox{i}}\left(g-j\right)\right)
d_{f,g}^+d_{f,g}
\end{eqnarray}
transforming (\ref{2.01}) into
\begin{eqnarray}
\label{3.02}
H
=\frac{1}{2}\sum_{i=0}^{N_x-1}\sum_{j=0}^{N_y-1}
\left(
J\left(d^+_{i,j}
{\mbox{e}}^{{\mbox{i}}\left(\alpha_{i+1,j}-\alpha_{i,j}\right)}
d_{i+1,j}
+d_{i,j}
{\mbox{e}}^{-{\mbox{i}}\left(\alpha_{i+1,j}-\alpha_{i,j}\right)}
d^+_{i+1,j}
\right.
\right.
\nonumber\\
\left.
\left.
+\gamma
\left(d^+_{i,j}
{\mbox{e}}^{-{\mbox{i}}\left(\alpha_{i+1,j}+\alpha_{i,j}\right)}
d^+_{i+1,j}
+d_{i,j}
{\mbox{e}}^{{\mbox{i}}\left(\alpha_{i+1,j}+\alpha_{i,j}\right)}
d_{i+1,j}\right)
\right)
\right.
\nonumber\\
\left.
+
J_{\perp}\left(
d^+_{i,j}
{\mbox{e}}^{{\mbox{i}}\left(\alpha_{i,j+1}-\alpha_{i,j}\right)}
d_{i,j+1}
+
d_{i,j}
{\mbox{e}}^{-{\mbox{i}}\left(\alpha_{i,j+1}-\alpha_{i,j}\right)}
d^+_{i,j+1}
\right.
\right.
\nonumber\\
\left.
\left.
+\gamma
\left(d^+_{i,j}
{\mbox{e}}^{-{\mbox{i}}\left(\alpha_{i,j+1}+\alpha_{i,j}\right)}
d^+_{i,j+1}
+
d_{i,j}
{\mbox{e}}^{{\mbox{i}}\left(\alpha_{i,j+1}+\alpha_{i,j}\right)}
d_{i,j+1}\right)\right)
\right).
\end{eqnarray}
We adopt a mean--field treatment of the phase factors
which have appeared in the Hamiltonian (\ref{3.02})
replacing the fermion occupation--number operators 
by their average value $\frac{1}{2}$ 
and using the same gauge as in Ref. \cite{005}
(see also Refs. \cite{004,009}).
As a result we arrive at
\begin{eqnarray}
\label{3.03}
H=\frac{1}{2}\sum_{i=0}^{N_x-1}\sum_{j=0}^{N_y-1}
\left(
J\left(-1\right)^{i+j}
\left(d^+_{i,j}d_{i+1,j}
-d_{i,j} d^+_{i+1,j}
+\gamma
\left(d^+_{i,j}d^+_{i+1,j}
-d_{i,j}d_{i+1,j}\right)
\right)
\right.
\nonumber\\
\left.
+
J_{\perp}\left(
d^+_{i,j} d_{i,j+1}
-d_{i,j} d^+_{i,j+1}
+\gamma
\left(d^+_{i,j} d^+_{i,j+1}
-
d_{i,j} d_{i,j+1}\right)\right)
\right).
\end{eqnarray}
This is the only approximate step made
to derive the thermodynamic quantities of the spin model (\ref{2.01})
within the 2D Jordan--Wigner fermionization approach.
Further consideration requires
no approximations.
Next we perform the Fourier transformation
\begin{eqnarray}
\label{3.04}
d_{i,j}=\frac{1}{\sqrt{N_xN_y}}
\sum_{k_x,k_y}
{\mbox{e}}^{{\mbox{i}}
\left(k_xi+k_yj\right)}
d_{k_x,k_y},
\;\;\;
d^+_{i,j}=\frac{1}{\sqrt{N_xN_y}}
\sum_{k_x,k_y}
{\mbox{e}}^{-{\mbox{i}}
\left(k_xi+k_yj\right)}
d^+_{k_x,k_y}
\end{eqnarray}
with
$k_x=\frac{2\pi}{N_x}n_x$,
$n_x=-\frac{N_x}{2},-\frac{N_x}{2}+1,\ldots,\frac{N_x}{2}-1$,
$k_y=\frac{2\pi}{N_y}n_y$,
$n_y=-\frac{N_y}{2},-\frac{N_y}{2}+1,\ldots,\frac{N_y}{2}-1$
($N_x$, $N_y$ are even).
As a result Eq. (\ref{3.03}) becomes
\begin{eqnarray}
\label{3.05}
H=\frac{1}{2}\sum_{{\bf{k}}}\nolimits^{\prime}
\left(
\begin{array}{cccc}
d^+_{k_x,k_y} &
d_{-k_x,-k_y} &
d^+_{k_x\pm\pi,k_y\pm\pi} &
d_{-k_x\pm\pi,-k_y\pm\pi}
\end{array}
\right)
\nonumber\\
\times
\left(
\begin{array}{cccc}
A & {\mbox{i}}B & -{\mbox{i}}C & -D\\
-{\mbox{i}}B & -A & D & {\mbox{i}}C \\
{\mbox{i}}C & D & -A & -{\mbox{i}}B \\
-D & -{\mbox{i}}C & {\mbox{i}}B & A
\end{array}
\right)
\left(
\begin{array}{c}
d_{k_x,k_y} \\
d^+_{-k_x,-k_y} \\
d_{k_x\pm\pi,k_y\pm\pi} \\
d^+_{-k_x\pm\pi,-k_y\pm\pi}
\end{array}
\right),
\nonumber\\
A=J_{\perp}\cos k_y,\;\;\;
B=\gamma J_{\perp}\sin k_y,\;\;\;
C=J\sin k_x,\;\;\;
D=\gamma J\cos k_x.
\end{eqnarray}
Here the prime denotes
that ${\bf{k}}$ in the thermodynamic limit varies in the region
$-\pi\le k_y\le\pi$,
$-\pi+\vert k_y\vert\le k_x\le\pi-\vert k_y\vert$.
The $4\times 4$ matrix in (\ref{3.05}) can be diagonalized.
Its eigenvalues
$\Lambda_{\alpha}({\bf{k}})$
are given by
\begin{eqnarray}
\label{3.06}
\Lambda_1({\bf{k}})
=\sqrt{\left(J_{\perp}\cos k_y+\gamma J\cos k_x\right)^2
+\left(J\sin k_x+\gamma J_{\perp}\sin k_y\right)^2},
\nonumber\\
\Lambda_2({\bf{k}})
=\sqrt{\left(J_{\perp}\cos k_y-\gamma J\cos k_x\right)^2
+\left(J\sin k_x-\gamma J_{\perp}\sin k_y\right)^2},
\nonumber\\
\Lambda_3({\bf{k}})=-\Lambda_1({\bf{k}}),
\nonumber\\
\Lambda_4({\bf{k}})=-\Lambda_2({\bf{k}}).
\end{eqnarray}
Therefore,
the Hamiltonian (\ref{3.05}) then assumes the form
\begin{eqnarray}
\label{3.07}
H=\sum_{{\bf{k}}}\nolimits^{\prime}
\sum_{\alpha=1}^2
\Lambda_{\alpha}({\bf{k}})
\left(\eta^+_{{\bf{k}},\alpha}\eta_{{\bf{k}},\alpha}
-\frac{1}{2}\right),
\end{eqnarray}
where
$\eta_{{\bf{k}},\alpha}$,
$\eta^+_{{\bf{k}},\alpha}$,
$\alpha=1,2$
are Fermi operators.

It is easy to calculate now
the Helmholtz free energy per site
\begin{eqnarray}
\label{3.08}
f=-\frac{1}{2\beta}
\int_{-\pi}^{\pi}\frac{{\mbox{d}}k_x}{2\pi}
\int_{-\pi}^{\pi}\frac{{\mbox{d}}k_y}{2\pi}
\left(
\ln\left(2\cosh\frac{\beta\Lambda_1({\bf{k}})}{2}\right)
+
\ln\left(2\cosh\frac{\beta\Lambda_2({\bf{k}})}{2}\right)
\right).
\end{eqnarray}
The ground--state energy per site
which follows from (\ref{3.08}) 
in the zero--temperature limit $\beta\to\infty$
reads
\begin{eqnarray}
\label{3.09}
e_0=
-\frac{1}{4}
\int
\frac{{\mbox{d}}{\bf{k}}}{\left(2\pi\right)^2}
\left(
\Lambda_1({\bf{k}})+\Lambda_2({\bf{k}})
\right).
\end{eqnarray}
The entropy and specific heat according to Eq. (\ref{3.08})
are as follows
\begin{eqnarray}
\label{3.10}
\frac{s}{k}
=\frac{1}{2}\int
\frac{{\mbox{d}}{\bf{k}}}{\left(2\pi\right)^2}
\left(
\ln\left(2\cosh\frac{\beta\Lambda_1({\bf{k}})}{2}\right)
+
\ln\left(2\cosh\frac{\beta\Lambda_2({\bf{k}})}{2}\right)
\right)
\nonumber\\
-\frac{1}{2}\int
\frac{{\mbox{d}}{\bf{k}}}{\left(2\pi\right)^2}
\left(
\frac{\beta\Lambda_1({\bf{k}})}{2}
\tanh\frac{\beta\Lambda_1({\bf{k}})}{2}
+
\frac{\beta\Lambda_2({\bf{k}})}{2}
\tanh\frac{\beta\Lambda_2({\bf{k}})}{2}
\right),
\end{eqnarray}
\begin{eqnarray}
\label{3.11}
\frac{c}{k}
=\frac{1}{2}\int
\frac{{\mbox{d}}{\bf{k}}}{\left(2\pi\right)^2}
\left(
\left(
\frac{\frac{\beta\Lambda_1({\bf{k}})}{2}}
{\cosh\frac{\beta\Lambda_1({\bf{k}})}{2}}
\right)^2
+
\left(
\frac{\frac{\beta\Lambda_2({\bf{k}})}{2}}
{\cosh\frac{\beta\Lambda_2({\bf{k}})}{2}}
\right)^2
\right).
\end{eqnarray}
In 1D limit formulas (\ref{3.06}), (\ref{3.08}) -- (\ref{3.11})
yield the long--known exact results
for the thermodynamic quantities
of spin--$\frac{1}{2}$ anisotropic $XY$ chain.
Note,
that the symmetry with respect to the change $\gamma$ to $-\gamma$
inherent in the original spin model
is present in the fermionic description (\ref{3.07}), (\ref{3.06}).

We cannot calculate the magnetization $m^x$ (or $m^y$)
by averaging $s_{i,j}^x$ (or $s_{i,j}^y$)
because of the invariance of the Hamiltonian (\ref{2.01})
when all $s^x_{i,j}\to -s^x_{i,j}$
(or all $s^y_{i,j}\to -s^y_{i,j}$);
this symmetry remains in the approximate treatment as well.
$m^x$ (or $m^y$)
can be obtained as a square root
of the two--site correlation function
of $x$ (or $y$) spin components
taken at sites which are at infinitely large distance.
However,
to our best knowledge
such correlation functions
have not been examined
within the 2D Jordan--Wigner fermionization approach
up till now.

In the second part of this Section
we consider the $XZ$ rather than the $XY$ Hamiltonian.
We start from the Hamiltonian (\ref{2.02}).
Making use of the transformation (\ref{3.01})
and adopting the mean--field treatment of phase factors
as used already for obtaining (\ref{3.03})
we get the following Hamiltonian
of interacting spinless fermions on a square lattice
\begin{eqnarray}
\label{3.12}
H
=\sum_{i=0}^{N_x-1}\sum_{j=0}^{N_y-1}
\left(
J\left(-1\right)^{i+j}\frac{1-\gamma}{4}
\left(d^+_{i,j} d^+_{i+1,j}+d^+_{i,j} d_{i+1,j}
-d_{i,j} d^+_{i+1,j}-d_{i,j} d_{i+1,j}\right)
\right.
\nonumber\\
\left.
+J\left(1+\gamma\right)
\left(d^+_{i,j} d_{i,j}-\frac{1}{2}\right)
\left(d^+_{i+1,j} d_{i+1,j}-\frac{1}{2}\right)
\right.
\nonumber\\
\left.
+
J_{\perp}\frac{1-\gamma}{4}
\left(d^+_{i,j} d^+_{i,j+1}+d^+_{i,j} d_{i,j+1}
-d_{i,j} d^+_{i,j+1}-d_{i,j} d_{i,j+1}\right)
\right.
\nonumber\\
\left.
+J_{\perp}\left(1+\gamma\right)
\left(d^+_{i,j} d_{i,j}-\frac{1}{2}\right)
\left(d^+_{i,j+1} d_{i,j+1}-\frac{1}{2}\right)
\right).
\end{eqnarray}

We need further approximations because of the interaction terms
in Eq. (\ref{3.12}).
We may assume that
\begin{eqnarray}
\label{3.13}
d^+_{i,j} d_{i,j}d^+_{i+1,j} d_{i+1,j}
\to
m d^+_{i,j}d_{i,j}
+\frac{1}{2}d^+_{i,j}d_{i,j}
+m d^+_{i+1,j}d_{i+1,j}
+\frac{1}{2}d^+_{i+1,j}d_{i+1,j}
-\frac{1}{4}-m^2-m,
\nonumber\\
d^+_{i,j} d_{i,j}d^+_{i,j+1} d_{i,j+1}
\to
m d^+_{i,j}d_{i,j}
+\frac{1}{2}d^+_{i,j}d_{i,j}
+m d^+_{i,j+1}d_{i,j+1}
+\frac{1}{2}d^+_{i,j+1}d_{i,j+1}
-\frac{1}{4}-m^2-m,
\end{eqnarray}
i.e., (ferromagnetic) long--range order 
(the existence of which in two dimensions is proved rigorously 
for $\gamma=0$ at zero temperature
or
for $\gamma=1$ at nonzero temperature)
is imposed. 
The order parameter is $m=m^z$ 
and its nonzero value may be expected for positive $\gamma$.
For negative $\gamma$
one may apply the unitary transformation 
$s^x\to  s^z$,
$s^y\to  s^y$,
$s^z\to -s^x$
coming back to the model with positive $\gamma$.
Thus the thermodynamics 
of the model with positive $\gamma$ and $m=m^z$ 
is the same as 
of the model with negative $\gamma$ and $m=m^x$.
These symmetry arguments permit us 
to find the thermodynamic properties of the model (\ref{2.02}) 
also for $\gamma\le 0$ 
basing on the results calculated for $\gamma\ge 0$.

Performing further the Fourier transformation (\ref{3.04})
one gets
\begin{eqnarray}
\label{3.14}
H=\frac{1}{2}\sum_{{\bf{k}}}\nolimits^{\prime}
\left(
\begin{array}{cccc}
d^+_{k_x,k_y} &
d_{-k_x,-k_y} &
d^+_{k_x\pm\pi,k_y\pm\pi} &
d_{-k_x\pm\pi,-k_y\pm\pi}
\end{array}
\right)
\nonumber\\
\times
\left(
\begin{array}{cccc}
{\cal{A}}+{\cal{M}}& {\mbox{i}}{\cal{B}} &
-{\mbox{i}}{\cal{C}} & -{\cal{D}}\\
-{\mbox{i}}{\cal{B}} & -{\cal{A}}-{\cal{M}} &
{\cal{D}} & {\mbox{i}}{\cal{C}} \\
{\mbox{i}}{\cal{C}} & {\cal{D}} &
-{\cal{A}}+{\cal{M}} & -{\mbox{i}}{\cal{B}} \\
-{\cal{D}} & -{\mbox{i}}{\cal{C}} &
{\mbox{i}}{\cal{B}} & {\cal{A}}-{\cal{M}}
\end{array}
\right)
\left(
\begin{array}{c}
d_{k_x,k_y} \\
d^+_{-k_x,-k_y} \\
d_{k_x\pm\pi,k_y\pm\pi} \\
d^+_{-k_x\pm\pi,-k_y\pm\pi}
\end{array}
\right)
-N\left(1+\gamma\right)\left(J+J_{\perp}\right)m^2
\end{eqnarray}
with
\begin{eqnarray}
\label{3.15}
{\cal{A}}=\frac{1-\gamma}{2}J_{\perp}\cos k_y,\;\;\;
{\cal{B}}=\frac{1-\gamma}{2}J_{\perp}\sin k_y,\;\;\;
{\cal{C}}=\frac{1-\gamma}{2}J\sin k_x,\;\;\;
{\cal{D}}=\frac{1-\gamma}{2}J\cos k_x,\;\;\;
\nonumber\\
{\cal{M}}=2\left(1+\gamma\right)\left(J+J_{\perp}\right)m.
\end{eqnarray}
The $4\times 4$ matrix in (\ref{3.14})
can be diagonalized.
Its eigenvalues $\Lambda_{\alpha}({\bf{k}})$ are given by
\begin{eqnarray}
\label{3.16}
\Lambda_{1,2}({\bf{k}})
=\sqrt{{\cal{A}}^2+{\cal{B}}^2+{\cal{C}}^2+{\cal{D}}^2+{\cal{M}}^2
\pm 2\sqrt{\left({\cal{A}}{\cal{D}}+{\cal{B}}{\cal{C}}\right)^2
+\left({\cal{A}}^2+{\cal{C}}^2\right){\cal{M}}^2}},
\nonumber\\
\Lambda_{3,4}({\bf{k}})=-\Lambda_{1,2}({\bf{k}}).
\end{eqnarray}
Therefore,
the resulting fermionic Hamiltonian assumes the form
\begin{eqnarray}
\label{3.17}
H={\sum_{\bf{k}}}^{\prime}\sum_{\alpha=1}^2
\Lambda_{\alpha}({\bf{k}})
\left(\eta^+_{{\bf{k}},\alpha}\eta_{{\bf{k}},{\alpha}}-\frac{1}{2}\right)
-N(1+\gamma)\left(J+J_{\perp}\right)m^2
\end{eqnarray}
with $\Lambda_{\alpha}({\bf{k}})$ (\ref{3.16}).

The Helmholtz free energy per site
is given by
\begin{eqnarray}
\label{3.18}
f=-\frac{1}{2\beta}
\int\frac{{\mbox{d}}{\bf{k}}}{\left(2\pi\right)^2}
\left(
\ln\left(2\cosh\frac{\beta\Lambda_1({\bf{k}})}{2}\right)
+
\ln\left(2\cosh\frac{\beta\Lambda_2({\bf{k}})}{2}\right)
\right)
+(1+\gamma)\left(\vert J\vert +\vert J_{\perp}\vert \right)m^2
\end{eqnarray}
and the magnetization $m$ is determined by minimizing (\ref{3.18}),
i.e., from the following equation
\begin{eqnarray}
\label{3.19}
2(1+\gamma)\left(\vert J\vert+\vert J_{\perp}\vert\right)m
=\frac{1}{4}
\int\frac{{\mbox{d}}{\bf{k}}}{\left(2\pi\right)^2}
\left(
\frac{\partial \Lambda_1({\bf{k}})}{\partial m}
\tanh\frac{\beta\Lambda_1({\bf{k}})}{2}
+
\frac{\partial \Lambda_2({\bf{k}})}{\partial m}
\tanh\frac{\beta\Lambda_2({\bf{k}})}{2}
\right).
\end{eqnarray}

As can be seen from (\ref{3.18}), (\ref{3.19}), (\ref{3.16}), (\ref{3.15})
the symmetry with respect to the change $\gamma$ to $-\gamma$
inherent in the spin model (\ref{2.02})
is destroyed in the elaborated formulas.
Evidently,
the derived results for the $XZ$ Hamiltonian  
in the limit $\gamma=-1$
coincide with the results
for the $XY$ Hamiltonian with $\gamma=\pm 1$
presented in the first part of this Section.

It can be easily observed 
that in the limit $\gamma=1$
the described 2D Jordan--Wigner fermionization approach
for the square--lattice $s=\frac{1}{2}$ $XZ$ model yields
the mean--field theory of the square--lattice Ising model.
Really, Eq. (\ref{3.12}) for $\gamma=1$
(in this case the fermionic representation (\ref{3.12}) becomes exact
since 
the Ising interaction of $z$ spin components 
does not involve the phase factors)
after decoupling
(\ref{3.13})
becomes
\begin{eqnarray}
\label{3.20}
H
=\sum_{i=0}^{N_x-1}\sum_{j=0}^{N_y-1}
\left(
2 J
\left(d^+_{i,j} d_{i,j}-\frac{1}{2}\right)
\left(d^+_{i+1,j} d_{i+1,j}-\frac{1}{2}\right)
+2 J_{\perp}
\left(d^+_{i,j} d_{i,j}-\frac{1}{2}\right)
\left(d^+_{i,j+1} d_{i,j+1}-\frac{1}{2}\right)
\right)
\nonumber\\
=\sum_{i=0}^{N_x-1}\sum_{j=0}^{N_y-1}
\left(
2 J
\left(m s^z_{i,j} +m s^z_{i+1,j}-m^2\right)
+2 J_{\perp}
\left(m s^z_{i,j} +m s^z_{i,j+1}-m^2\right)
\right)
\end{eqnarray}
(we have used the relation
$s^z=d^+d-\frac{1}{2}$).
Obviously,
the Helmholtz free energy per site 
for the mean--field spin Hamiltonian (\ref{3.20})
follows also from
(\ref{3.18})
after inserting $\gamma=1$,
i.e.,
$f=-\frac{1}{\beta}
\ln\left(2\cosh\left(2\beta\left(\vert J\vert+\vert J_{\perp}\vert\right)m\right)\right)
+2\left(\vert J\vert+\vert J_{\perp}\vert\right)m^2$.
The equation for $m$ (\ref{3.19}) becomes simple
\begin{eqnarray}
\label{3.21}
m=\frac{1}{2}\tanh
\left(2\beta\left(\vert J\vert+\vert J_{\perp}\vert\right)m\right)
\end{eqnarray}
and hence we arrive at familiar formulas
for the internal energy per site
$e=-2\left(\vert J\vert+\vert J_{\perp}\vert\right)m^2$
and the specific heat
$\frac{c}{k}
=4\beta^2\left(\vert J\vert+\vert J_{\perp}\vert\right)
m\frac{\partial m}{\partial\beta}$.

To close this Section
let us note,
that a more sophisticated treatment
(i.e., a more complicated decoupling of the l.h.s. of (\ref{3.13}))
of the Hamiltonian (\ref{3.12})
is possible
(see Ref. \cite{009}
for a review of analogous studies for the isotropic Heisenberg model).
However,
we found 
that this more complicated approximation 
seems to be less adequate
to describe the properties of the spin system.

\section{Finite--size scaling analysis of
the exact diagonalization data}

\setcounter{equation}{0}

In this Section we present the results
of the exact diagonalization computations
of the ground--state energy
and the ground--state magnetization
for the $s=\frac{1}{2}$ isotropic (i.e., $\gamma=0$) $XY$ model
on small spatially anisotropic (i.e., $0\le R\le 1$) lattices
with periodic boundary conditions.
The results 
for
$e_0$
and
${m^x}^2=\frac{1}{N_x^2N_y^2}\sum_{i_1,i_2=1}^{N_x}\sum_{j_1,j_2=1}^{N_y}
\langle s_{i_1,j_1}^xs_{i_2,j_2}^x\rangle$
for spatially anisotropic finite square lattices
of $N=N_xN_y$ sites
($N_x=N_y=\sqrt{N}=L$)
for $L=4$ and $L=6$
of square shape 
are reported in Table \ref{t1}.
\begin{table}
\begin{center}

\caption
{\small 
Exact diagonalization data
for the spatially anisotropic $s=\frac{1}{2}$ $XY$ model
($\gamma=0$) on finite square lattices
of square shape.}

\vspace{5mm}

\begin{tabular}{|c|c|c|c|c|}                                            \hline
$R$  & $e_0(L=4)$   & $e_0(L=6)$   & ${m^x}^2(L=4)$ & ${m^x}^2(L=6)$ \\ \hline
0.00 & $-0.353553$  & $-0.333333$  &  0.045535      &  0.025463      \\ \hline
0.01 & $-0.353586$  & $-0.333385$  &  0.047178      &  0.027241      \\ \hline
0.02 & $-0.353684$  & $-0.333541$  &  0.048908      &  0.029217      \\ \hline
0.03 & $-0.353848$  & $-0.333801$  &  0.050731      &  0.031427      \\ \hline
0.04 & $-0.354078$  & $-0.334164$  &  0.052652      &  0.033911      \\ \hline
0.05 & $-0.354375$  & $-0.334632$  &  0.054673      &  0.036713      \\ \hline
0.10 & $-0.356909$  & $-0.338570$  &  0.066284      &  0.056225      \\ \hline
0.15 & $-0.361291$  & $-0.345166$  &  0.079674      &  0.078399      \\ \hline
0.20 & $-0.367541$  & $-0.353772$  &  0.092803      &  0.092460      \\ \hline
0.30 & $-0.384698$  & $-0.374145$  &  0.111863      &  0.105124      \\ \hline
0.40 & $-0.405918$  & $-0.396792$  &  0.121758      &  0.110963      \\ \hline
0.50 & $-0.429449$  & $-0.420846$  &  0.126876      &  0.114331      \\ \hline
0.60 & $-0.454432$  & $-0.445916$  &  0.129709      &  0.116406      \\ \hline
0.70 & $-0.480432$  & $-0.471769$  &  0.131334      &  0.117683      \\ \hline
0.80 & $-0.507203$  & $-0.498253$  &  0.132243      &  0.118429      \\ \hline
0.90 & $-0.534590$  & $-0.525255$  &  0.132689      &  0.118804      \\ \hline
1.00 & $-0.562486$  & $-0.552694$  &  0.132816      &  0.118911      \\ \hline
\end{tabular}
\label{t1}
\end{center}
\end{table}
We also calculate $e_0$ and  
${m^x}^2=\frac{1}{N^2}\sum_{m_1=1}^{N}\sum_{m_2=1}^{N}
\langle s_{m_1}^xs_{m_2}^x\rangle$
for spatially anisotropic finite square lattices 
of $N=8,\;18,\;32$ sites 
of diamond shape
(Table \ref{t2}).
\begin{table}
\begin{center}

\caption
{\small 
Exact diagonalization data
for the spatially anisotropic $s=\frac{1}{2}$ $XY$ model
($\gamma=0$) on finite square lattices 
of diamond shape.}

\vspace{5mm}

\begin{tabular}{|c|c|c|c|c|c|c|}                             \hline
$R$  & $e_0(N=8)$   & $e_0(N=18)$     &  $e_0(N=32)$     &
     ${m^x}^2(N=8)$ & ${m^x}^2(N=18)$ &  ${m^x}^2(N=32)$  \\ \hline
0.00 & $-0.353553$  & $-0.333333$     &  $-0.326641$     &
        0.091069    &  0.050926       &   0.033695        \\ \hline
0.01 & $-0.353617$  & $-0.333387$     &  $-0.326712$     &
        0.094262    &  0.054471       &   0.037441        \\ \hline
0.02 & $-0.353809$  & $-0.333555$     &  $-0.326929$     &
        0.097420    &  0.058335       &   0.041851        \\ \hline
0.03 & $-0.354127$  & $-0.333845$     &  $-0.327298$     &
        0.100523    &  0.062478       &   0.046982        \\ \hline
0.04 & $-0.354570$  & $-0.334266$     &  $-0.327828$     &
        0.103553    &  0.066833       &   0.052763        \\ \hline
0.05 & $-0.355134$  & $-0.334823$     &  $-0.328526$     &
        0.106492    &  0.071311       &   0.058967        \\ \hline
0.10 & $-0.359648$  & $-0.339627$     &  $-0.334342$     &
        0.119459    &  0.092111       &   0.085177        \\ \hline
0.15 & $-0.366554$  & $-0.347155$     &  $-0.342674$     &
        0.129211    &  0.105816       &   0.097591        \\ \hline
0.20 & $-0.375259$  & $-0.356377$     &  $-0.352257$     &
        0.136131    &  0.113676       &   0.103995        \\ \hline
0.30 & $-0.396250$  & $-0.377584$     &  $-0.373529$     &
        0.144342    &  0.121607       &   0.111009        \\ \hline
0.40 & $-0.420165$  & $-0.400902$     &  $-0.396613$     &
        0.148511    &  0.125525       &   0.114920        \\ \hline
0.50 & $-0.445817$  & $-0.425543$     &  $-0.420942$     &
        0.150783    &  0.127801       &   0.117335        \\ \hline
0.60 & $-0.472609$  & $-0.451135$     &  $-0.446204$     &
        0.152087    &  0.129200       &   0.118862        \\ \hline
0.70 & $-0.500218$  & $-0.477457$     &  $-0.472195$     &
        0.152846    &  0.130060       &   0.119815        \\ \hline
0.80 & $-0.528446$  & $-0.504366$     &  $-0.498777$     &
        0.153274    &  0.130561       &   0.120375        \\ \hline
0.90 & $-0.557169$  & $-0.531756$     &  $-0.525847$     &
        0.153484    &  0.130813       &   0.120657        \\ \hline
1.00 & $-0.586302$  & $-0.559552$     &  $-0.553329$     &
        0.153543    &  0.130885       &   0.120738        \\ \hline
\end{tabular}
\label{t2}
\end{center}
\end{table}
It is worth to remark
that according to the adopted definition of ${m^x}^2$ 
both the results 
for the square and diamond shaped finite square lattices
for $R=0$ 
yield the adequate value for a chain 
divided by the number of isolated chains 
into which the finite square lattice splits for $R=0$
(compare, for example, the results 
from Table \ref{t1} and Table \ref{t2}
for ${m^x}^2(L=4)$ 
(4 chains at $R=0$)
and  ${m^x}^2(N=8)$
(2 chains at $R=0$)
or 
for ${m^x}^2(L=6)$ 
(6 chains at $R=0$)
and  ${m^x}^2(N=18)$
(3 chains at $R=0$)).

From \cite{034,022,023} we know
that the ground--state energy per site
for the case $R=1$, $\gamma=0$
scales as
\begin{eqnarray}
\label{4.01}
e_0(L)=e_0(\infty)
+\frac{\epsilon_3}{L^3}
+\frac{\epsilon_5}{L^5}
+\ldots.
\end{eqnarray}
From Ref. \cite{037}
we know that the finite--size corrections to the ground--state energy
per site
for the case $R=0$, $\gamma=0$
(for both periodic and antiperiodic boundary conditions)
scales as
\begin{eqnarray}
\label{4.02}
e_0(N_x)-e_0(\infty)
\sim\frac{1}{{N_x}^2}.
\end{eqnarray}
To extrapolate the exact diagonalization data
with periodic boundary conditions
for $0\le R\le 1$
we use the scaling law 
\begin{eqnarray}
\label{4.04}
e_0(L)=e_0(\infty)
+\frac{\epsilon}{L^{2+R}}
\end{eqnarray}
which contains two unknown parameters $e_0(\infty)$ and $\epsilon$
and yields correct asymptotics in 1D and 2D limits. 
The same scaling law was used in \cite{013} 
to extrapolate the exact diagonalization data 
for spatially anisotropic 2D $s=\frac{1}{2}$ Heisenberg model.
Another scaling law which also yields correct asymptotics 
in 1D and 2D limits
and therefore may be assumed
to extrapolate the exact diagonalization data
for $0\le R\le 1$,
\begin{eqnarray}
\label{4.05}
e_0(L)
=e_0(\infty)
+\frac{\epsilon_2}{L^2}
+\frac{\epsilon_3}{L^3}
+\ldots,
\end{eqnarray}
contains three parameters
$e_0(\infty)$, $\epsilon_2$, $\epsilon_3$
to be determined from the exact diagonalization data.
Bearing in mind the 2D--to--1D crossover  
we should note, 
that the available data for $N=16,\;36$ square lattices
cannot be completed by data for $N=25$ square lattice
since the latter lattice for $R\to 0$
transforms into chains of odd number of sites
and the data essentially differ 
from those for chains of even numbers of sites.
The exact diagonalization data for $N=64$ 
are far beyond the available computer resources.
On the other hand,
the lattices of $N=8,\;18,\;32$ sites
in the limit $R\to 0$
split into 2, 3, 4  chains of 4, 6, 8 sites
and the shape effects become more pronounced.

The simplest assumption
for extrapolation of the exact diagonalization data
for ${m^x}^2(L)$
for $R=1$, $\gamma=0$ 
(see, for example, \cite{023,038})
is as follows
\begin{eqnarray}
\label{4.06}
{m^x}^2\left(L\right)
={m^x}^2(\infty)
+\frac{\mu_1}{L}
+\ldots.
\end{eqnarray}
We use
the finite--size scaling (\ref{4.06}) 
to extrapolate the numerical data
for $N=16,\;36$
and 
for $N=8,\;18,\;32$
lattices for $0\le R\le 1$.
The finite--size scaling analysis
of the available data yields
incorrect prediction ${m^x}^2(\infty)<0$
in quasi--1D limit starting from $R$ 
in between 0.03 and 0.05 
(see Section 5 below)
that may serve as indication of a disappearance of the long--range order
in 1D limit.

From \cite{039} we know that in the 1D case $R=0$, $\gamma=0$ 
$x$ (or $y$) magnetization vanishes as 
\begin{eqnarray}
\label{4.06a}
{m^x}^2\left(N_x\right)
\sim\frac{1}{\sqrt{N_x}}.
\end{eqnarray}
To obtain ${m^x}^2(\infty)$ for $0\le R\le 1$
we may combine (\ref{4.06a}) and (\ref{4.06})
and assume ${m^x}^2\left(L\right)$
to behave as follows 
\begin{eqnarray}
\label{4.06b}
{m^x}^2\left(L\right)
={m^x}^2(\infty)
+\frac{\mu}{L^{\frac{1+R}{2}}}
\end{eqnarray}
or as follows 
\begin{eqnarray}
\label{4.07}
{m^x}^2\left(L\right)
={m^x}^2(\infty)
+\frac{\mu_{\frac{1}{2}}}{\sqrt{L}}
+\frac{\mu_1}{L}
+\ldots.
\end{eqnarray}
The $\frac{1}{\sqrt{L}}$ scaling 
which appears in (\ref{4.06b}) and (\ref{4.07}) as $R\to 0$ 
(as well as in (\ref{4.06a}) for $R=0$)
when applied to accessible small systems
works poorly
probably because $L$ is still too small.

To summarize,
using the data for $N=16,\;36$ and $N=8,\;18,\;32$ clusters
and scaling laws (\ref{4.04}) and (\ref{4.06}) 
we find $e_0(\infty)$ and ${m^x}^2(\infty)$
(and the order parameter
$m=\sqrt{{m^x}^2+{m^y}^2}=\sqrt{2{m^x}^2}$).
We use these findings for the discussion in the next Section.
Let us also remark,
that the presented discourse refers to the case $\gamma=0$,
whereas the case of anisotropic exchange interaction $\gamma\ne 0$
was not considered.

\section{Effects of the spatial and exchange interaction
anisotropies}

\setcounter{equation}{0}

In this Section we discuss the results 
for the ground--state energy, the order parameter and the specific heat 
obtained by the different methods illustrated in  Sections 2, 3, 4.
We also compare the approximate and exact results when available.

\begin{figure}
\epsfysize=200mm
\epsfclipon
\centerline{\epsffile{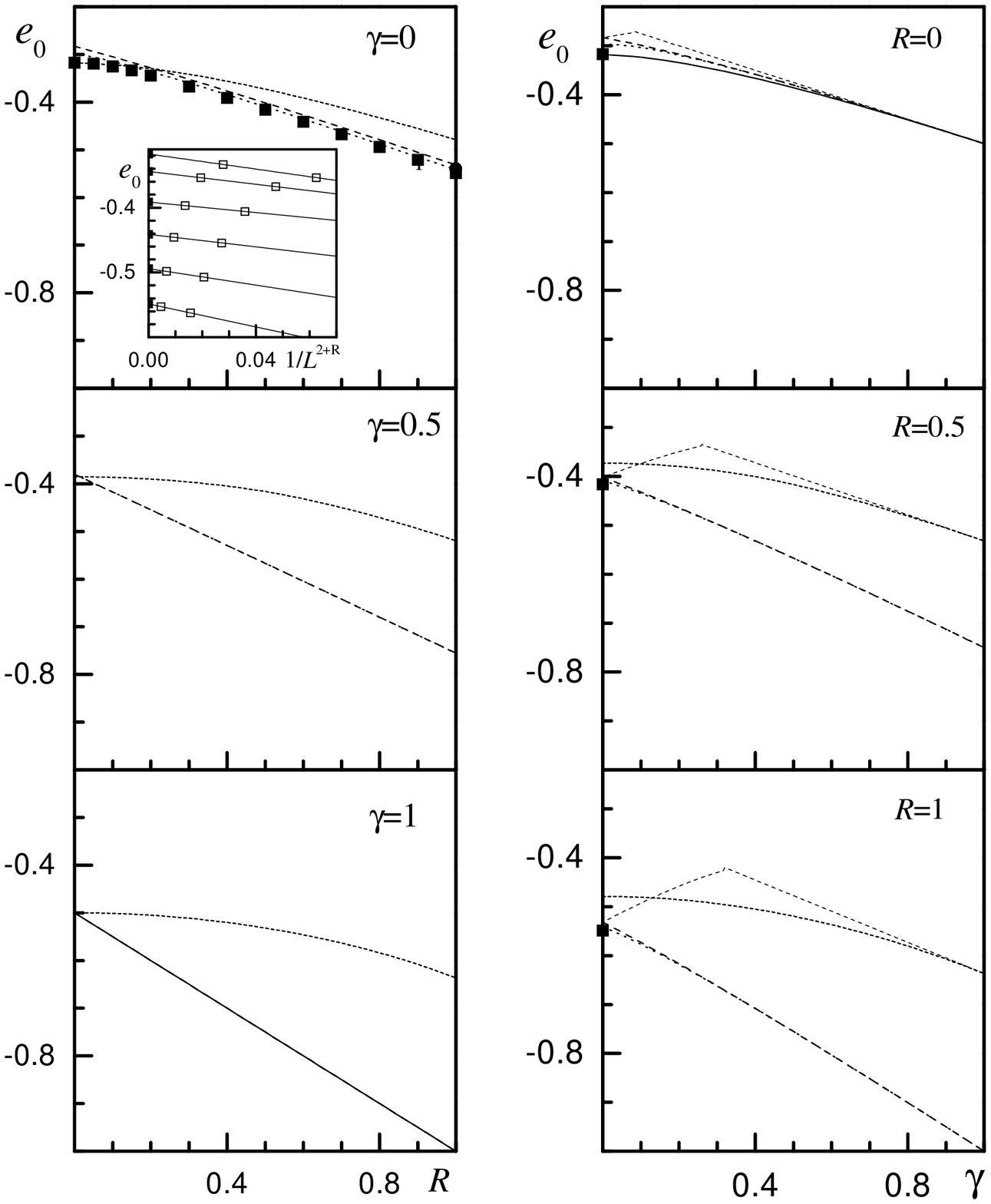}}
\caption[]
{\small
The ground--state energy 
of the square--lattice 
$s=\frac{1}{2}$ $XY$ ($XZ$) model:
exact results (solid curves),
finite--size extrapolation ($N=16,\;36$) (full squares),
linear spin--wave theory (dotted curves),
fermionization approach for $XY$ model (short--dashed curves)
and $XZ$ model (long--dashed curves).
Thin long--dashed curves with cusps in the right panels
correspond to the dependence $e_0$ vs. $-\gamma$
as it follows from (\ref{3.18}), (\ref{3.19}) 
in the limit $\beta\to\infty$.
The inset demonstrates size scaling (\ref{4.04})
of the finite lattice data
($N=16,\;36$)
for $R=1,0.8,0.6,0.4,0.2, 0$
(from bottom to top).}
\label{fig001}
\end{figure}

In Fig. \ref{fig001}
we show the ground--state energy
of the square--lattice $s=\frac{1}{2}$ $XY$ model
obtained within various approaches
in comparison with some exact results
(the 1D limit and the Ising limit).
As can be seen from Eq. (\ref{3.09}), (\ref{3.06})
the Jordan--Wigner fermions
for the $XY$ model (short--dashed curves)
recover the 1D limit ($R=0$, $R=\infty$).
In the other limit of the Ising model ($\gamma=1$)
this approach fails (obviously except the 1D limit $R=0$, $R=\infty$).
The Jordan--Wigner fermions for $XZ$ model
(long--dashed curves)
correctly reproduce the dependence $e_0$ vs. $R$ for $\gamma=1$
but not the exact results for $R=0$.
We also show the dependence $e_0$ vs. $-\gamma$
obtained within the latter approximation
(thin long--dashed curves with cusps)
to demonstrate explicitly a difference in the outcomes
as $\gamma$ changes its sign.
It should be remarked
that the linear spin--wave theory predictions
for the ground--state energy (\ref{2.09})
(dotted curves)
are in a satisfactory agreement
with the exact results for $R=0$ (except small $\gamma$),
coincide with the exact result for $\gamma=1$
and are in a good agreement
with the exact diagonalization data for $\gamma=0$
(full squares).

\begin{figure}
\epsfysize=80mm
\epsfclipon
\centerline{\epsffile{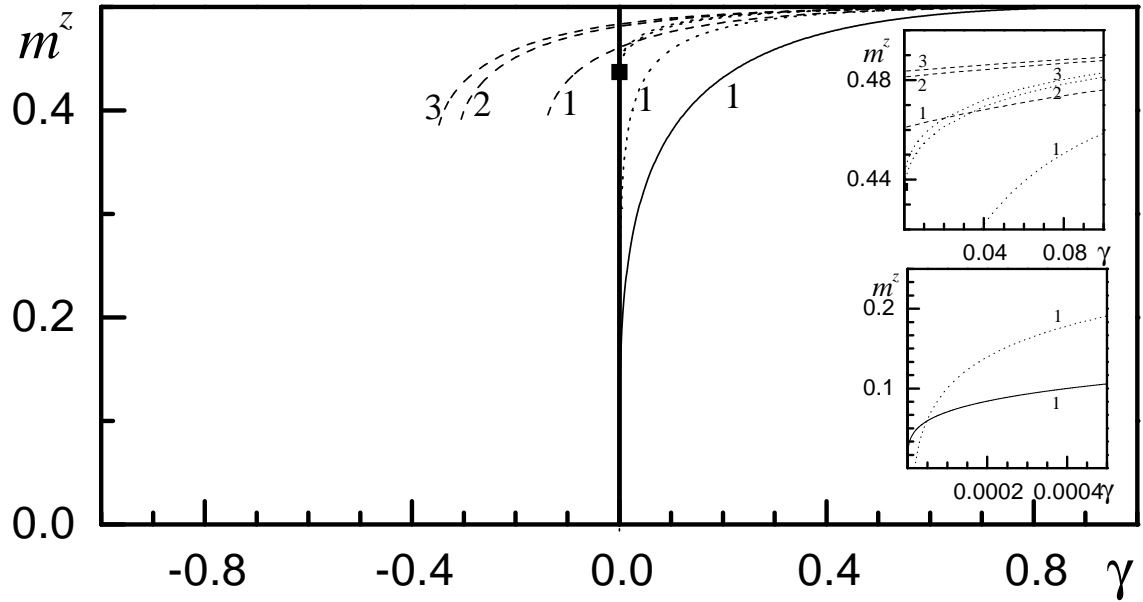}}
\caption[]
{\small
The zero--temperature magnetization $m^z$
vs. 
the exchange interaction anisotropy parameter $\gamma$ 
for square--lattice $s=\frac{1}{2}$ $XZ$ model
(1: $R=0$,
2: $R=0.5$,
3: $R=1$):
exact result for $R=0$ (solid curves),
finite--size extrapolation result ($N=16,\;36$)
for $R=1$ 
(full square),
linear spin--wave theory results (dotted curves)
and fermionization approach results (long--dashed curves).
In the lower inset the spin--wave theory result for $R=0$
becomes negative for $\gamma$ less than $1.7\cdot 10^{-5}$.}
\label{fig002}
\end{figure}

The $XZ$ model 
within the fermionic picture
exhibits in--plane magnetization
introduced by decoupling (\ref{3.13}),
which is determined
self--consistently by Eq. (\ref{3.19}).
Fig. \ref{fig002} shows
the resulting dependence of the ground--state magnetization
$m^z$
(long--dashed curves)
on $\gamma$
for $R=1$ (curves 3), $R=0.5$ (curves 2), and $R=0$ (curves 1).
The value of $m^z$ for $R=1$
overestimates the exact diagonalization result at $\gamma=0$
(full square).
Moreover,
the magnetization does not disappear in 1D limit at $\gamma=0$
that contradicts the rigorous statement
and indicates the incorrect treatment 
of strong quantum fluctuations in one dimension.
The exact zero--temperature relation 
between in--plane magnetization 
and exchange anisotropy parameter
in 1D reads
$m
=\frac{1}{2}
\left(1-\left(\frac{1-\gamma}{1+\gamma}\right)^2\right)^{\frac{1}{4}}$
\cite{039}
(solid curves 1 in Fig. \ref{fig002}).
The linear spin--wave theory results (\ref{2.10})
(dotted curves)
are valid only for $\gamma\ge 0$
as was explained in Section 2.
(Therefore,
within the frames of the linear spin--wave theory
we cannot trace the behaviour of $m^z$
(for example, for $R=1$)
when decreasing $\gamma$ to negative values.)
Contrary to the fermionic description
the linear spin--wave theory predicts smaller values of $m^z$ at $\gamma=0$
in a good agreement
with the exact diagonalization data
(see the upper inset in Fig. \ref{fig002}).
The linear spin--wave theory value of $m^z$ for $\gamma=0$
decreases as $R$ decreases
and finally becomes zero at a certain small but nonzero value of $R$.
If $R\to 0$ 
the value of $m^z$ for $\gamma=0$ tends to $-\infty$
which is a manifestation of the inapplicability
of the linear spin--wave theory in the 1D limit
(see the lower inset in Fig. \ref{fig002}
where the spin--wave theory result for $R=0$ is shown).

\begin{figure}
\epsfysize=80mm
\epsfclipon
\centerline{\epsffile{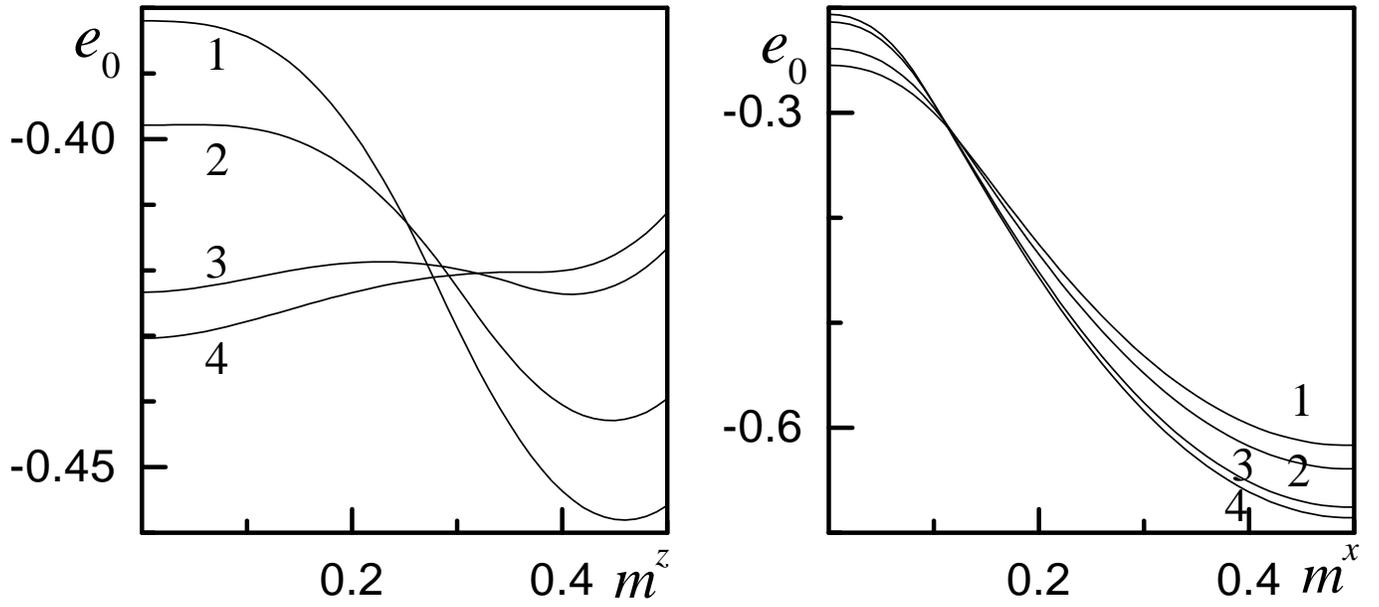}}
\caption[]
{\small
The dependence of
the ground--state energy of spatially isotropic ($R=1$)
$s=\frac{1}{2}$ $XZ$ model
on the order parameter 
directed along $z$ (left panel) or $x$ (right panel) axis
(obtained from (\ref{3.18}) as $\beta\to\infty$)
(1: $\gamma=-0.2$,
2: $\gamma=-0.25$,
3: $\gamma=-0.33$,
4: $\gamma=-0.352$).}
\label{fig003}
\end{figure}

\begin{figure}
\epsfysize=80mm
\epsfclipon
\centerline{\epsffile{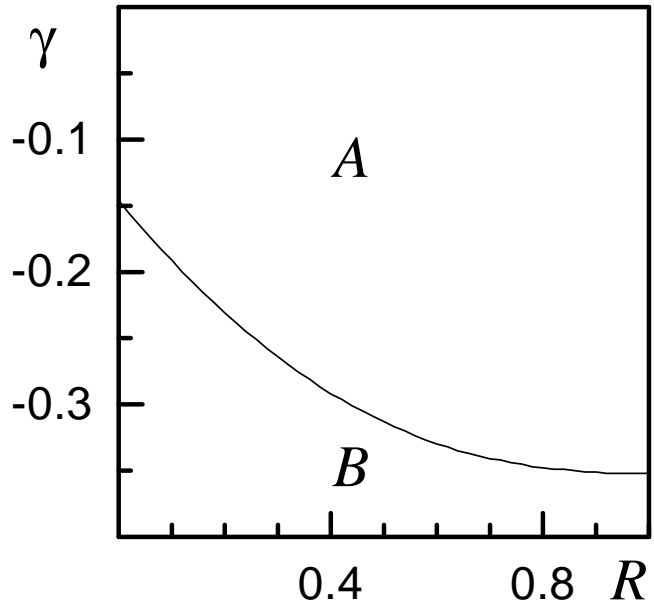}}
\caption[]
{\small
Region of metastability 
for the square--lattice $s=\frac{1}{2}$ $XZ$ model
as it follows from the fermionization approach.
For all $\gamma <0$ the stable phase has $m^z=0$, $m^x\ne 0$.
In the region $A$ 
the phase with $m^z\ne 0$ is still possible 
although not favourable (metastable phase),
in region $B$ the phase with $m^z\ne 0$ can not exist.
The solid curve 
which separates the regions $A$ and $B$ 
represents the dependence $\gamma_3$ vs. $R$
(see explanation in the main text).}
\label{fig004}
\end{figure}

Let us consider the dependence 
of the ground--state in--plane magnetization on $\gamma$
as it issues from the fermionic picture
in more detail.
Assume, for example, the spatially isotropic case $R=1$.
From exact results for this model
we know
that for $\gamma=1$
the order parameter $m=\frac{1}{2}$
is directed along $z$ axis,
whereas for $\gamma=+0$
the order parameter is $0<m<\frac{1}{2}$
and it remains directed along $z$ axis.
If $\gamma$ 
becomes negative (tending to $-1$)
two possibilities are plausible:
i)
for infinitesimally small $\gamma=-0$
the order parameter $m$ turns to the $x$ axis
and its value starts to increase
(approaching $\frac{1}{2}$ as $\gamma\to -1$);
ii)
the order parameter $m$ remains directed along $z$ axis
and its value decreases
until a certain characteristic value of (negative) $\gamma$
at which $m$ abruptly turns to $x$ axis
and then its value starts to increase
(approaching $\frac{1}{2}$ as $\gamma\to -1$).
The fermionization approach results
give evidence in favour of the latter scenario.
To illustrate this issue
we show in Fig. \ref{fig003} the ground--state energy
(given by (\ref{3.18}) as $\beta\to\infty$)
for spatially isotropic ($R=1$) $XZ$ model 
as a function of $m^z$ or $m^x$ 
(see the comments after Eq. (\ref{3.13}))
for several values of $\gamma<0$.
The profiles plotted in the left panel 
exhibit
one minimum at $m^z\ne 0$ for $\gamma_1<\gamma<0$
(curve 1),
two minima at $m^z\ne 0$ (the deeper one) and at $m^z=0$ for $\gamma_2<\gamma<\gamma_1$
(curve 2),
two minima at $m^z\ne 0$ and at $m^z=0$ of the same depth for $\gamma=\gamma_2$
(curve 3),
two minima at $m^z\ne 0$ and at $m^z=0$ (the deeper one) for $\gamma_3<\gamma<\gamma_2$
(curve 4),
and
one minimum at $m^z=0$ for $\gamma_3<\gamma$.
(The curve $m^z$ vs. $\gamma$ in Fig. \ref{fig002} 
jumps to zero value at $\gamma=\gamma_3$.)
The profiles plotted in the right panel
represent the dependence of the ground--state energy 
for the same $\gamma$s
on the order parameter directed along $x$ axis.
From these plots we see that for $\gamma <0$ 
the minimum in the dependence $e_0$ vs. $m^x$ 
is always deeper 
than the one in the dependence $e_0$ vs. $m^z$
and hence the $x$ directed order parameter is favourable.
However,
while decreasing $\gamma$ from positive values to negative ones 
the system may remain in the metastable phase 
with the order parameter directed along $z$ axis 
until $\gamma$ achieves $\gamma_3$. 
This is a typical scenario of the first--order quantum phase transition
driven by the exchange interaction anisotropy $\gamma$.
The corresponding region of metastability 
in the $\gamma-R$--plane
is shown in Fig. \ref{fig004}.
The obtained issue can be compared
with the predictions of other approaches.
The linear spin--wave theory does not work if $\gamma<0$
and thus cannot be used for this discussion.
The coupled cluster method
and
the correlated basis function method
similarly to the 2D Jordan--Wigner fermionization approach
suggest an analogous behaviour of the ground--state energy
\cite{025,026}.
Moreover,
the correlated basis function prediction
for the value of $\gamma=-0.36$
associated with a phase transition
for spatially isotropic square--lattice \cite{026}
is quite close to the estimate obtained by the fermionization approach
$\gamma_3=-0.3528$.
An intriguing question naturally arises here:
is the described scenario
of a first--order quantum phase transition 
inherent in the spin model
or it is an artifact of the approximate approaches?
This question, apparently, cannot be answered by the mentioned theories
and more work is necessary to draw a definite conclusion.

\begin{figure}
\epsfysize=80mm
\epsfclipon
\centerline{\epsffile{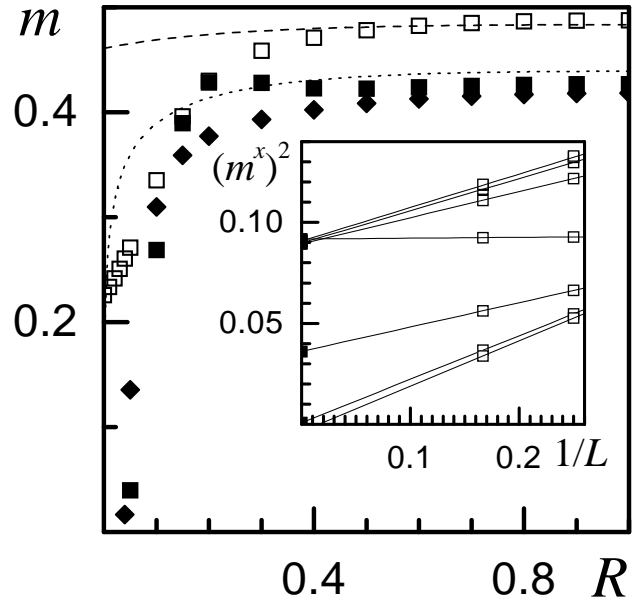}}
\caption[]
{\small
The order parameter for the square--lattice $s=\frac{1}{2}$
isotropic (i.e., $\gamma=0$) $XY$ ($XZ$) model
as the spatial anisotropy parameter $R$ varies from 0 to 1:
exact diagonalization data 
for $N=36$ square shaped clusters
(empty squares in the main plot),
exact diagonalization data completed by scaling (\ref{4.06})
(square shaped clusters: full squares,
diamond shaped clusters: full diamonds),
linear spin--wave theory (dotted curve),
fermionization approach (long--dashed curve).
The linear spin--wave theory result for $m$
becomes zero at $R$ about $7\cdot 10^{-4}$.
The inset demonstrates the size scaling (\ref{4.06}) 
of square shaped clusters 
for $R=0.04,0.05,0.1,0.2,0.4,0.6,0.8,1$ (from bottom to top;
the curves for $R=0.8$ and $R=1$ coincide in the chosen scale).}
\label{fig005}
\end{figure}

Next we pass to another interesting problem,
that is,
the appearance/disappearance of the long--range order
at zero temperature in quasi--1D limit.
It is known that there is no long--range order
for the 1D ($R=0$) isotropic ($\gamma=0$) $XY$ model
and it does exist
for the spatially isotropic 2D ($R=1$) isotropic ($\gamma=0$) $XY$ model.
If $\gamma=1$ the order parameter takes its classical value 
$m=\frac{1}{2}$ for all $0\le R\le 1$.
Let us consider the dependence $m$ vs. $R$
which arises within different approaches.
The prediction for the order parameter $m^z$ vs. $R$
as it follows from the Jordan--Wigner picture
for the $XZ$ model with $\gamma=0$ 
is shown in Fig. \ref{fig005} by long--dashed curve.
The exact result $m^z=0$ for $R=0$ 
is not reproduced.
The result of the linear spin--wave theory
is shown in Fig. \ref{fig005} by dotted curve.
The linear spin--wave theory does not work properly as $R \to 0$
yielding $m^z\to -\infty$.
To make this explicit,
we insert (\ref{2.07}) into (\ref{2.10})
and note
that the integrand diverges at $R=0$, $\gamma=0$ around $k_x=0$.
We expand the integrand about small $k_x$
to pick up the main contribution to the integral
in the limit $R,\gamma\to 0$.
For $\gamma=0$, $R\to 0$ we get
\begin{eqnarray}
\label{5.01}
m^z=\frac{1}{4\sqrt{2}\pi}\ln R+{\mbox{finite terms}}.
\end{eqnarray}
For finite $\gamma>0$
the order parameter $m^z$ is finite for all $0\le R\le 1$;
if $R=0$
the order parameter 
again has a logarithmic divergence as $\gamma\to 0$,
$m^z=\frac{1}{4\sqrt{2}\pi}\ln\gamma+{\mbox{finite terms}}$;
for finite $R>0$
the order parameter $m^z$ is finite for all $0\le \gamma\le 1$.
The dotted curve in Fig. \ref{fig005} for very small $R$
reproduces the logarithmic divergence for $\gamma=0$
and the characteristic value of $R$
at which $m^z$ becomes zero
is about $7\cdot 10^{-4}$.
We also report  
the exact diagonalization data 
for the $N=36$ lattice 
shown by empty squares in the main plot in Fig. \ref{fig005}
as well as
the finite--size scaling (\ref{4.06}) predictions
from square and diamond shaped clusters   
shown by full squares and full diamonds,
respectively.
According to (\ref{4.06})
${m^x(\infty)}^2$ becomes negative 
for $R$ 
in between 
0.03 and 0.04 (diamond shaped clusters)
or 
in between 
0.04 and 0.05 (square shaped clusters)
(see, for example, 
two lower curves in the inset in Fig. \ref{fig005}
which refer to the square shaped clusters)
that we may take as a criterion
for disappearance of the long--range order
as the spin system is becoming a collection of noninteracting chains.

We may summarize our findings as follows:
for the spatially anisotropic 2D $s=\frac{1}{2}$ isotropic 
(i.e., $\gamma=0$) 
$XY$ ($XZ$) model
the 2D Jordan--Wigner fermionization approach
predicts long--range order for any $0\le R\le 1$,
the linear spin--wave theory for $R>R_c$ with $R_c$ about $7\cdot 10^{-4}$,
the finite--size scaling data (\ref{4.06})
for $R>R_c$ with $R_c$ in between 0.03 and 0.05.
It should be remarked
that the results of the performed analysis
are qualitatively similar
to what have been obtained for weakly coupled
$s=\frac{1}{2}$ isotropic Heisenberg chains
\cite{011,012,013}
(see also a recent study
for a 2D array of interacting two--leg $s=\frac{1}{2}$ Heisenberg ladders
\cite{014}).
Note, however, 
that the exact diagonalization data prediction for $R_c$ 
for the $XY$ Hamiltonian 
is in between 0.03 and 0.05 
whereas for the Heisenberg Hamiltonian
$R_c$ is essentially larger  
been in between 0.2 and 0.3 
(see Fig. 2 of Ref. \cite{013}).
For both systems,
i.e.,
the isotropic $XY$ model
and
the isotropic Heisenberg model
the question about developing of the long--range order
as the interchain interaction starts to increase
is far from being settled
and more delicate approaches are necessary to resolve this question.

\begin{figure}
\epsfysize=200mm
\epsfclipon
\centerline{\epsffile{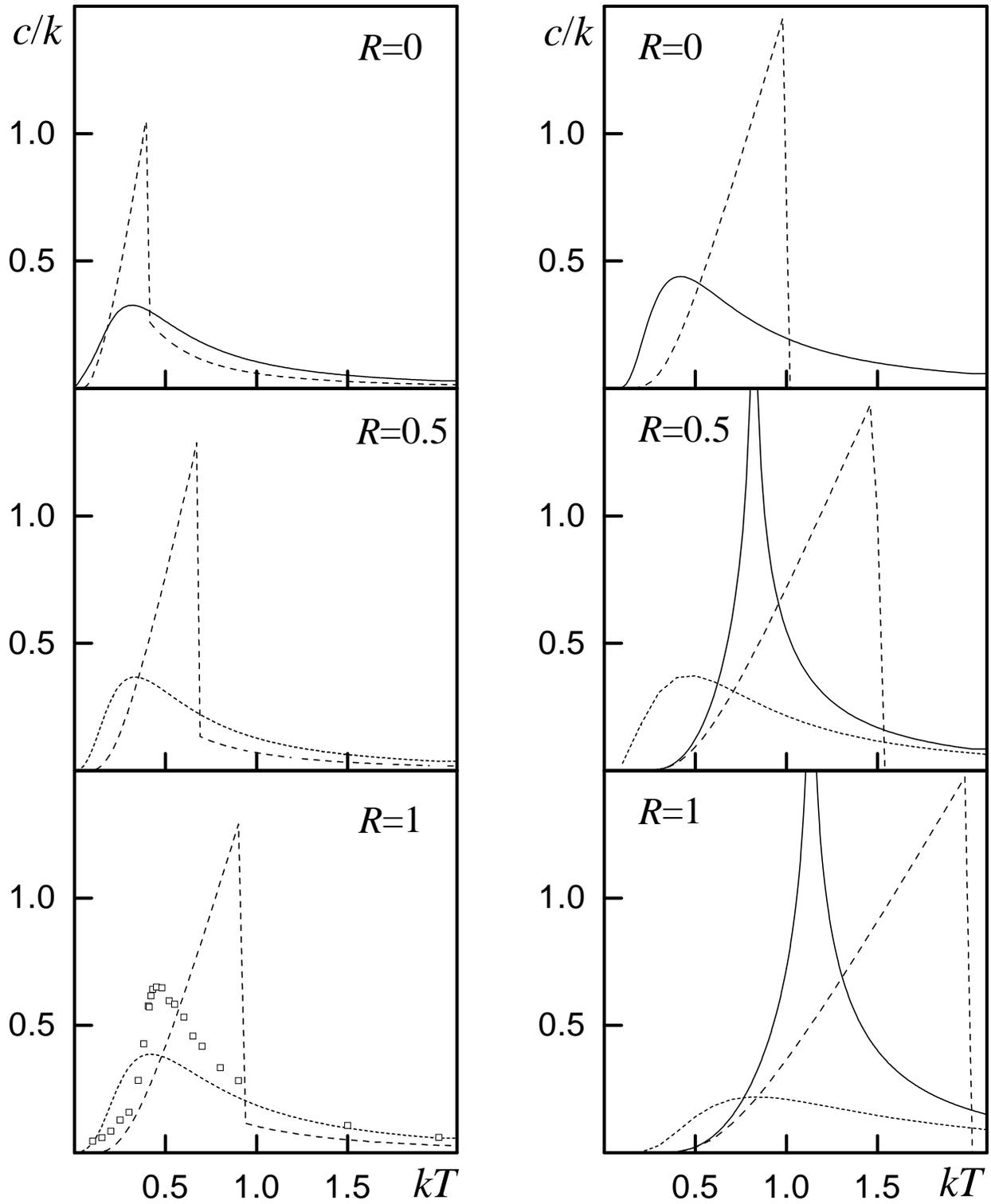}}
\caption[]
{\small
The temperature dependence of the specific heat
for the square--lattice $s=\frac{1}{2}$ $XY$ ($XZ$) model:
exact results (solid curves),
quantum Monte Carlo result (open squares) \cite{020},
fermionization approach results for $XY$ model
(short--dashed curves)
and for $XZ$ model (long--dashed curves).
The left panels refer to the case $\gamma=0$,
the right panels refer to the case $\gamma=1$.
For $R=0$ (two upper panels) 
the solid and short--dashed curves coincide.}
\label{fig006}
\end{figure}

Let us turn to the properties
of the considered square--lattice $s=\frac{1}{2}$ model
at nonzero temperatures.
In Fig. \ref{fig006} we show
the temperature dependence of the specific heat
for the
$XY$ model (\ref{3.11}) (short--dashed curves)
and
$XZ$ model (which follows from (\ref{3.18}), (\ref{3.19})) (long--dashed curves)
in comparison with
the exact results (solid curves) for $R=0$
and
the Onsager solutions (for $\gamma=1$) \cite{040}
as well as with
the quantum Monte Carlo result
for $\gamma=0$, $R=1$ \cite{020} (open squares).
Eq. (\ref{3.11}) yields exact result in 1D limit.
The noninteracting Jordan--Wigner fermions (\ref{3.07}), (\ref{3.06})
cannot reproduce the logarithmic singularity inherent in the spin model
for $\gamma=1$
(however, recover the exact result
as the square--lattice Ising model becomes a set of noninteracting Ising chains).
The temperature dependence of the specific heat
for $XZ$ model as it follows from (\ref{3.18}), (\ref{3.19})
has a mean--field character
exhibiting an incorrect finite jump.
A noticeable difference between the outcomes
of the quantum Monte Carlo
and 2D Jordan--Wigner fermionization approaches for $\gamma=0$, $R=1$
is seen at low and intermediate temperatures.
Recently the temperature dependence of the specific heat
for $\gamma=0$, $R=1$ has been examined
within the Green's--function approach \cite{027}.
The obtained temperature profile
(Fig. 2 of that paper)
although is in qualitative agreement with the quantum Monte Carlo data
but exhibits a lower peak at a higher temperature.

\begin{figure}
\epsfysize=70mm
\epsfclipon
\centerline{\epsffile{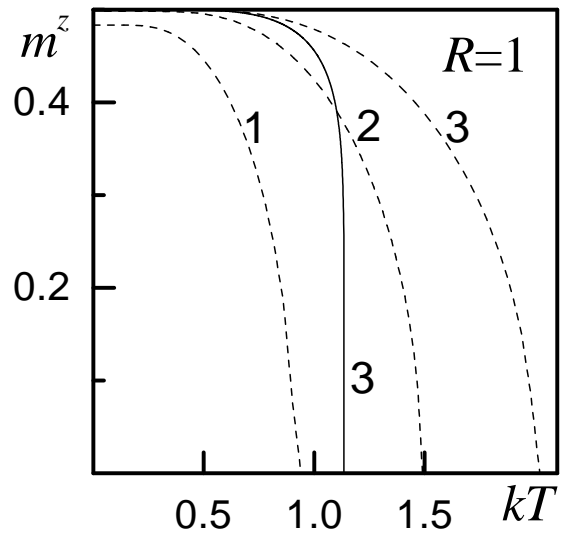}}
\caption[]
{\small
The temperature dependence of the order parameter $m^z$
for spatially isotropic ($R=1$) $s=\frac{1}{2}$ $XZ$ model
(1: $\gamma=0$,
2: $\gamma=0.5$,
3: $\gamma=1$):
exact (solid curve)
and fermionization approach (long--dashed curves) results.}
\label{fig007}
\end{figure}

In Fig. \ref{fig007} we plot
the temperature dependence of the in--plane magnetization
for spatially isotropic 2D ($R=1$) $XZ$ model 
derived within the 2D Jordan--Wigner fermionization scheme
(long--dashed curves)
and the exact result for $\gamma=1$
(solid curve).
For $R=1$, $\gamma=1$ the approximate result
based on the fermionization approach
(long--dashed curve 3 in Fig. \ref{fig007})
is the mean--field counterpart of the Onsager solution
(solid curve 3 in Fig. \ref{fig007}).
The approximate result for $R=1$, $\gamma=0$
(long--dashed curve 1 in Fig. \ref{fig007})
is incorrect 
since the Mermin--Wagner theorem predicts
$m=0$ at any finite temperature
for the isotropic $XY$ model.
For $R=0$ 
the fermionization approach
yields decreasing of $m^z$ as temperature increases 
and vanishing of $m^z$ at a certain temperature 
instead of the correct behaviour $m=0$ for any finite temperature.

\section{Summary}

To summarize,
we have extended the 2D Jordan--Wigner fermionization approach
for the {\it anisotropic} square--lattice $s=\frac{1}{2}$ $XY$ model.
In our study of the ground--state properties
we have concentrated on the effects of
i) exchange interaction anisotropy
and
ii) spatial anisotropy.
We have analyzed
1) the quantum phase transition
driven by the exchange interaction anisotropy
and
2) appearance/disappearance of the long--range order in quasi--1D limit.
We have discussed these issues
using different approaches:
2D Jordan--Wigner fermionization,
linear spin--wave theory
and exact diagonalization data supplemented by finite--size scaling.
The study of the anisotropy driven changes extends the 
the earlier papers \cite{025,026} 
suggesting the first--order quantum phase transition scenario.
The study of the 2D--to--1D crossover 
constitutes the $XY$ counterpart of the problem
intensively discussed within the context of the Heisenberg model 
(\cite{013} and references therein). 
In our study of the temperature effects
we have discussed
to what extent the Jordan--Wigner fermions
can reproduce the peculiarities
of the specific heat
for the square--lattice Ising and isotropic $XY$ models
as well as 
the relation to 
the Mermin--Wagner theorem.

We have found that the 2D Jordan--Wigner fermionization
for the $XY$ Hamiltonian works well in quasi--1D limit
(especially as $\gamma\to 0$)
for calculating both
the ground--state and thermodynamic quantities.
Being applied to the $XZ$ Hamiltonian
the 2D Jordan--Wigner fermionization gives the ground--state results similar
to the linear spin--wave theory.
Although both approaches reproduce the existing long--range order 
in two dimensions 
(and therefore work well as $\gamma\to 1$),
they overestimate its effects
and fail as $R\to 0$.
The temperature effects for the $XZ$ model are poorly reproduced
within the 2D Jordan--Wigner fermionization scheme.

\section*{Acknowledgments}

The present study was partly supported by the DFG
(projects 436 UKR 17/7/01 and  436 UKR 17/1/02). 
O. D. acknowledges the kind hospitality of the Magdeburg University
in the summer of 2001 and in the summer of 2002.
O. D. thanks the STCU for the support (project \#1673).
The paper was partly presented at
the XII School of Modern Physics
on Phase Transitions and Critical Phenomena
(L\c{a}dek Zdr\'{o}j, 21 -- 24 June 2001)
and
the 11--th Czech and Slovak Conference on Magnetism
(Ko\v{s}ice, 20 -- 23 August 2001).
O. D. thanks the organizers of both meetings
and
T. V. thanks the organizers of the latter meeting
for the support for participation.

\clearpage

\clearpage

\clearpage

\clearpage

\clearpage

\clearpage

\clearpage

\end{document}